\documentclass[10pt,conference]{IEEEtran}
\usepackage{amsmath}

 \DeclareMathOperator{\tr}{tr}

\newcommand{\F}{\mathbf{F}}

\newcommand{\nix}[1]{}
\newcommand{\onemat}{\mathbf{1}}
\newcommand{\mbf}{\mathbf}

\DeclareMathOperator{\wt}{wt}

\newcommand{\RM}{{\mathcal{R}}}

\newcommand{\hdual}{{\bot_h}}

\newtheorem{theorem}{Theorem}
\newtheorem{lemma}{Lemma}
\newtheorem{corollary}[lemma]{Corollary}

\begin{document}
\title{Nonbinary Quantum Reed-Muller Codes}
\author{\authorblockN{Pradeep K. Sarvepalli}
\authorblockA{Dept. of Computer Science\\
Texas A\&M University\\
College Station, TX 77843-3112, USA \\
Email: pradeep@cs.tamu.edu} \and
\authorblockN{Andreas Klappenecker}
\authorblockA{Dept. of Computer Science\\
Texas A\&M University\\
College Station, TX 77843-3112, USA \\
Email: klappi@cs.tamu.edu} }
%

\maketitle

\begin{abstract}
We construct nonbinary quantum codes from classical generalized
Reed-Muller codes and derive the conditions under which these quantum
codes can be punctured.  We provide a partial answer to a question
raised by Grassl, Beth and R{\"o}tteler on the existence of $q$-ary
quantum MDS codes of length $n$ with $q\le n\le q^2-1$. 
\end{abstract}

\section{Introduction}
The quest to build a scalable quantum computer that is resilient
against decoherence errors and operational noise has sparked a lot of
interest in quantum error-correcting codes. The early research has
been confined to the study of binary quantum error-correcting codes,
but more recently the theory was extended to nonbinary codes that are
useful in the realization of fault-tolerant computations. 

The purpose of this paper is to derive families of quantum stabilizer
codes that are based on classical generalized Reed-Muller (GRM) codes.
Recall that one cannot arbitrarily shorten a quantum
stabilizer code. We study the puncture codes of GRM stabilizer codes
that determine the possible lengths of shortened codes.

We give now an overview of our main theorems. We omit some technical
details to keep our exposition brief, but we assure the reader that
the subsequent sections provide all missing details.  Let $q$ be power
of a prime. We denote by $\RM_q(\nu,m)$ the $q$-ary generalized
Reed-Muller code of order $\nu$ with parameters
$[q^m,k(\nu),d(\nu)]_q$ and dual distance $d(\nu^\perp)$.

Our first two results concern the construction of two families of
nonbinary quantum stabilizer codes:
\smallskip

\begin{theorem}\label{th:grm_css1}
For $0\leq \nu_1 \leq \nu_2 \leq m(q-1)-1$, there exist pure
quantum stabilizer codes with the parameters
$[[q^m,k(\nu_2)-k(\nu_1),\min\{d(\nu_1^\perp),d(\nu_2)\}]]_q.$
\end{theorem}
\smallskip

\begin{theorem}\label{th:grm_herm}
For $0 \leq \nu \leq m(q-1)-1$, there exists a pure
$[[q^{2m},q^{2m}-2k(\nu), d(\nu^\perp)]]_q$ quantum stabilizer code.
\end{theorem}
\smallskip

Puncturing a quantum stabilizer code is restricted because the
underlying classical code must remain self-orthogonal. Our next two
results show when quantum Reed-Muller codes can be punctured.
\smallskip

\begin{theorem}\label{th:grm_punc1}
For $0\leq \nu_1 \leq \nu_2 \leq m(q-1)-1$ and $0\leq \mu\leq
\nu_2-\nu_1$, if $\mathcal{R}_q(\mu,m)$ has a codeword of weight
$r$, then there exists an $[[r,\geq (k(\nu_2)-k(\nu_1)-q^m+r),\geq d]]_q$
quantum stabilizer code with $d=\min\{d(\nu_2),d(\nu_1^\perp)\}$.
\end{theorem}
\smallskip

\begin{theorem}\label{th:grm_punc2}
Let $C=\mathcal{R}_{q^2}(\nu,m)$ with $0\leq \nu\leq m(q-1)-1$ and
$ (q+1)\nu \leq \mu \leq m(q^2-1)-1$. If
$\mathcal{R}_{q^2}(\mu,m)^\perp|_{\F_q}$ contains a vector of
weight $r$, then there exists a quantum stabilizer code with parameters 
$[[r,\geq (r-2k(\nu)),\geq d(\nu^\perp)]]_q$.
\end{theorem}
\smallskip

Our last result deals with the existence of quantum MDS codes of
length $n$ with $q\le n\le q^2-1$. These are derived from the previous
results on punctured codes making use of some additional properties of
MDS codes.
\smallskip

\begin{theorem}\label{th:mds_lengths}
There exist quantum MDS codes with the parameters $[[(\nu+1)q,
(\nu+1)q-2\nu-2,\nu+2]]_q$ for $0\leq \nu\leq q-2$.
\end{theorem}
\smallskip

The paper is organized as follows. In the next section we review the
basics of generalized Reed-Muller codes. We construct two series of
quantum codes using a nonbinary version of the CSS construction and a
Hermitian construction. In Section~\ref{sec:punc}, we derive our
results concerning the puncturing of these stabilizer codes.
In
Section~\ref{sec:mds}, we consider a special case of GRM codes to
obtain some quantum MDS codes.  The MDS nature of these codes allows
us to make some more definitive statements about their puncture codes,
making it possible to show the existence of quantum MDS codes of
lengths between $q$ and $q^2$.
\smallskip

\paragraph*{Notation} We denote  the Euclidean
inner product of two vectors $x$ and $y $ in $\F_q^n$ by $\langle x |
y\rangle=x_1y_1+\cdots x_ny_n$. We write $\langle x |
y\rangle_h=x_1y_1^q+\cdots x_ny_n^q$ to denote the Hermitian inner
product of two vectors $x$ and $y$ in $\F_{q^2}^n$. 
We denote the 
Euclidean dual of a code $C\subseteq \F_q^n$ by $C^\perp$, and 
the Hermitian dual of a code $D\subseteq \F_{q^2}^n$ by $D^\hdual$.

\setcounter{theorem}{0} 
\section{Generalized Reed-Muller Codes}
Primitive generalized Reed-Muller codes were introduced by Kasami,
Lin, and Petersen~\cite{kasami68} as a generalization of Reed-Muller
codes~\cite{muller54,reed54}.  We follow Assmus and
Key~\cite{assmus92,assmus98} in our approach to GRM codes.  Our main
goal is to derive two series of quantum stabilizer codes.  

\subsection{Classical Codes}
 Let
$(P_1,\dots, P_n)$ denote an enumeration of all points in $\F_q^m$
with $n=q^m$. We denote by $L_m(\nu)$ the subspace of
$\F_q[x_1,\dots, x_m]$ that is generated by polynomials of degree
$\nu$ or less. Let $\nu$ denote an integer in the range $0\le \nu<
m(q-1)$.  Let $ev$ denote the evaluation function 
$ev$~$f =(f(P_1),\dots,f(P_n))$.  
The generalized Reed-Muller code $\RM_q(\nu,m)$ of order
$\nu$ is defined as
\begin{eqnarray}
 \RM_q(\nu,m) &=& \{ (f(P_1),\dots,f(P_n))\,|\, f\in L_m(\nu) \},\\
&=&\{ ev \mbox{ }f \,|\, f\in L_m(\nu) \}.\nonumber
\end{eqnarray}
The dimension $k(\nu)$ of the code $\RM_q(\nu,m)$ equals
\begin{eqnarray}k(\nu) &= &\sum_{j=0}^m (-1)^j
\binom{m}{j}\binom{m+\nu-jq}{\nu-jq}, \label{eq:grm_dim}
\end{eqnarray}
and its minimum distance $d(\nu)$ is given by
\begin{eqnarray}
d(\nu)&=&(R+1)q^Q, \label{eq:grm_dist}
\end{eqnarray}
where $ m(q-1)-\nu = (q-1)Q+ R$, such that $0\le R<q-1$; see
\cite{assmus92,assmus98,kasami68,pellikaan04}.

It is clear that $\RM_q(\nu,m)\subseteq \RM_q(\nu',m)$ holds for
all parameters $\nu\le \nu'$. More interesting is the fact that
the dual code of $\RM_q(\nu,m)$ is again a generalized Reed-Muller
code,
$$\RM_q(\nu,m)^\perp = \RM(\nu^\perp,m) \quad \text{with}
\quad \nu^\perp = m(q-1)-1-\nu.$$ We need the following result for
determining the distances and purity of quantum codes.
\smallskip

\begin{lemma}\label{th:grmwt}
Let $C_1=\mathcal{R}_q(\nu_1,m)$ and $C_2 = \mathcal{R}_q(\nu_2,m)$.
If $v_1<v_2$, then $C_1\subset C_2$ and $\wt(C_2\setminus C_1)=
\wt(C_2)$.
\end{lemma}
\smallskip 
\begin{proof}
We already know that $C_1\subset C_2$ if $\nu_1 <\nu_2$.
We denote the minimum distances of the codes $C_1$ and $C_2$ by
$d_1=\mbox{wt}(C_1)=(R_1+1)q^{Q_1}$ and $d_2=
\wt(C_2)=(R_2+1)q^{Q_2}$. 

It suffices to show that $\nu_1<\nu_2$ implies $d_2<d_1$, because in
that case $C_2\setminus C_1$ must contain a vector of weight $d_2$,
and this shows that $\wt(C_2\setminus C_1)=\wt(C_2)$, as claimed.

Since $\nu_1<\nu_2$, we have 
\begin{equation}\label{ineq} 
m(q-1)-\nu_2 < m(q-1)-\nu_1.
\end{equation}
If we set $m(q-1)-\nu_k = (q-1)Q_k+R_k$ with $0\le R_k<q-1$ for $k\in \{1,2\}$, 
then it follows from (\ref{ineq}) that $Q_1\ge Q_2$. 

If $Q_1 = Q_2$ then $R_1>R_2$; hence, 
$d_1=(R_1+1)q^{Q_1}>(R_2+1)q^{Q_2}=d_2.$
On the other hand, if $Q_1 > Q_2$, then 
$d_1\geq (0+1)q^{Q_1}$ and $d_2\leq (q-2+1)q^{Q_2}=(q-1)q^{Q_2}$, 
and it follows that $d_1\geq q^{Q_2+1} > d_2 $. 
\end{proof}
\subsection{Quantum codes}

Quantum Reed-Muller codes were first constructed by Steane \cite{steane99}
based on classical Reed-Muller codes. 
We derive a series of stabilizer codes from generalized
Reed-Muller codes using the CSS code construction  and a
Hermitian code construction.
\smallskip

\begin{lemma}\label{th:css}
Let $C_1=[n,k_1,d_1]_q$ and $C_2=[n,k_2,d_2]_q$ be linear codes
over $\F_q$ with $C_1 \subseteq C_2$. Furthermore, let
$d=\min \mbox{wt}\{(C_2\setminus C_1)\cup (C_1^\perp\setminus
C_2^\perp)\}$ if $C_1 \subset C_2$ and 
$d= \min\mbox{wt}\{ (C_1)\cup (C_1^\perp)\}$ 
 if $C_1=C_2$. Then there exists an
 $[[n,k_2-k_1,d]]_q$ quantum code.
\end{lemma}
\smallskip

\begin{proof}
See for instance \cite{calderbank98} for the
CSS construction of binary codes and  \cite[Theorem 3]{grassl03} 
and \cite{kim04} for its $q$-ary generalizations.
\end{proof}
\smallskip

\begin{theorem}
For $0\leq \nu_1 \leq \nu_2 \leq m(q-1)-1$, there exists a pure
$[[q^m,k(\nu_2)-k(\nu_1),\min\{d(\nu_1^\perp),d(\nu_2)\}]]_q$ quantum
stabilizer code, where the parameters $k(\nu)$ and $d(\nu)$ are given
by the equations (\ref{eq:grm_dim}) and (\ref{eq:grm_dist}), respectively.
\end{theorem}
\smallskip

\begin{proof}
For $\nu_1 \leq \nu_2$, $C_1=\mathcal{R}_q(\nu_1,m) \subseteq
\mathcal{R}_q(\nu_2,m)=C_2$.  By Lemma~\ref{th:css}, we know there
exists a pure $[[q^m,k(\nu_2)-k(\nu_1),\min\{d(\nu_1^\perp),d(\nu_2)\}]]_q$
quantum code. The purity of the code follows from Lemma~\ref{th:grmwt}.
\end{proof}
\smallskip

\begin{corollary} \label{co:grm_css2}
For  $0 \leq \nu \leq (m(q-1)-1)/2$ there exists a pure
$[[n,n-2k(\nu), d(\nu^\perp)]]_q$ quantum code, where $k(\nu)$ is
given by equation (\ref{eq:grm_dim}) and the distance $d(\nu)$ by
equation (\ref{eq:grm_dist}).
\end{corollary}
\smallskip

\begin{proof}
In Theorem \ref{th:grm_css1} if we choose $\nu_1=\nu$,
$\nu_2=\nu^\perp$, then to ensure $\nu_1\leq \nu_2$ we require $\nu
\leq (m(q-1)-1)/2$. That the distance of the resulting code is
$d(\nu^\perp)$ follows by direct substitution of these values in
Theorem \ref{th:grm_css1}.
\end{proof}
\smallskip

The next construction starts from a generalized Reed-Muller code
over $\F_{q^2}$. If it is contained in its Hermitian dual code, then
it can be used to construct quantum codes. 
Therefore, our immediate goal will be to find such self-orthogonal 
Reed-Muller codes.
\smallskip

\begin{lemma}\label{th:grm_hdual}
If $\nu$ is an order in the range $0\le \nu \leq m(q-1)-1$, then
$\RM_{q^2}(\nu,m)\subseteq \RM_{q^2}(\nu,m)^\hdual$.
\end{lemma}
\smallskip

\begin{proof}
Recall that $ev$~$f=(f(P_1),\dots,f(P_n))$.  The code $\RM_{q^2}(0,m)$ is
generated by $\onemat$, the all one vector. The relation
$\RM_{q^2}(0,m)^\perp = \RM_{q^2}(m(q^2-1)-1,m)$ shows that
$\langle ev$~$f | \onemat \rangle=0$ for all polynomials $f$ in $L_m(\nu)$
with $\deg f \le m(q^2-1)-1$. If $x^{a_1}_1\cdots x^{a_m}_m$ and
$ x^{b_1}_1\cdots x^{b_m}_m$ are monomials in $L_m(\nu)$, then
\[
\begin{split}
\langle ev\, x^{a_1}_1&\cdots x^{a_m}_m \,|\, 
ev\,x^{b_1}_1\cdots x^{b_m}_m \rangle_h\\{}
&=
\langle ev\, x^{a_1}_1\cdots x^{a_m}_m \,|\, 
ev\,x^{qb_1}_1\cdots  x^{qb_m}_m \rangle, \\
&= \langle ev\, x^{a_1+qb_1}_1\cdots x^{a_m+qb_m}_m \,|\, 
\onemat\rangle = 0,
\end{split}
\]
where the last equality holds because the monomial has at most
degree $(m(q-1)-1)(q+1) < m(q^2-1)-1$. Since the monomials
generate $L_m(\nu)$, it follows that $\langle ev$~$f |\, ev\,
$~$g \rangle_h=0$ for all $f, g$ in $L_m(\nu)$.
\end{proof}
\smallskip

The following construction, which we refer to as the Hermitian construction,
directly leads to the second series of quantum codes that can be 
constructed from the Reed-Muller codes.

\begin{lemma} \label{th:herm}
Let $C$ be a linear $[n,k]_{q^2}$ contained in its Hermitian dual,
$C^\hdual$, such that $d=\min\{\mbox{wt}(C^\hdual\setminus C)\}$.
Then there exists an $[[n,n-2k,d]]_q$ quantum code.
\end{lemma}
\smallskip

\begin{proof}
See for instance \cite[Corollary 1]{ashikhmin01} and 
\cite[Corollary 2]{grassl03}.
\end{proof}
\smallskip

\begin{theorem}
For $0 \leq \nu \leq m(q-1)-1$, there exist pure quantum codes
$[[q^{2m},q^{2m}-2k(\nu), d(\nu^\perp)]]_q$, where
$$
k(\nu) = \sum_{j=0}^m (-1)^j
\binom{m}{j}\binom{m+\nu-jq^2}{\nu-jq^2},
$$
and $d(\nu^\perp)=(R+1)q^{2Q}$, with $\nu+1 =(q^2-1)Q+R$ and
$0\leq R<q^2-1$.
\end{theorem}
\smallskip

\begin{proof}
First we note that wt$(\mathcal{R}_{q^2}(\nu,m)^\hdual)= $
wt$(\mathcal{R}_{q^2}(\nu,m)^\perp)=d(\nu^\perp)$. Recall that
$d(\nu^\perp)$  can be computed using equation (\ref{eq:grm_dist}),
keeping in mind that these codes
are over $\F_{q^2}$. From Lemma \ref{th:grm_hdual} and Lemma
\ref{th:herm} we can conclude that there exists a pure quantum
code $[[q^{2m},q^{2m}-2k(\nu), d(\nu^\perp)]]_q$ where 
$k(\nu)$ is the dimension of $\mathcal{R}_{q^2}(\nu,m)$ as given
by equation (\ref{eq:grm_dim}). The purity of the code follows
from Lemma \ref{th:grmwt}.
\end{proof}


\section{Puncturing Quantum GRM Codes}\label{sec:punc}
Puncturing provides a means to construct new codes from existing
codes.  Puncturing quantum stabilizer codes, however, is not as
straightforward as in the case of classical codes. Rains introduced
the notion of puncture code \cite{rains99}, which simplified this
problem and provided a means to find out when punctured codes are
possible.  Further extensions to these ideas can be found in
\cite{grassl03}.  With the help of these results we now study the
puncturing of GRM codes.

Recall that for every quantum code constructed using the CSS construction,
we can associate two classical codes, $C_1$ and $C_2$. 
Define $C$ to be the direct sum of
$C_1$ and $C_2^\perp$ viz. $C=C_1\oplus C_2^\perp$. The
puncture code $P(C)$ \cite[Theorem 12]{grassl03} is defined as
\begin{eqnarray}
P(C)&=&\{ (a_ib_i)_{i=1}^{n} \mid a \in C_1,b\in C_2^\perp\}^\perp.
\label{eq:puncdef1}
\end{eqnarray}
The usefulness of the puncture codes lies in the fact that if there
exists a vector of nonzero weight $r$, then the corresponding quantum
code can be punctured to a length $r$ and minimum distance greater
than or equal to distance of the parent code. 
\smallskip

\begin{theorem}
For $0\leq \nu_1 \leq \nu_2 \leq m(q-1)-1$ and $0\leq \mu\leq
\nu_2-\nu_1$, if $\mathcal{R}_q(\mu,m)$ has codeword of weight
$r$, then there exists an $[[r,\geq (k(\nu_2)-k(\nu_1)-q^m+r),\geq d]]_q$
quantum code, where $d=\min\{d(\nu_2),d(\nu_1^\perp)
\}$. In particular, there exists a
$[[d(\mu),\geq (k(\nu_2)-k(\nu_1)-q^m+d(\mu)),\geq d]]_q$ quantum code.
\end{theorem}
\smallskip

\begin{proof}
Let $C_i=\mathcal{R}_q(\nu_i,m)$ with $0\leq \nu_1 \leq \nu_2\leq
m(q-1)-1$, for $i \in \{1,2 \}$. By Theorem~\ref{th:grm_css1}, a 
$[[q^m,k(\nu_2)-k(\nu_1),d]]_q$ quantum code
$Q$ with $d=\min\{d(\nu_2),d(\nu_1^\perp)\}$ exists. 
It follows from equation (\ref{eq:puncdef1}) that 
$P(C)^\perp=\mathcal{R}_q(\nu_1+\nu_2^\perp,m)$, so
\begin{eqnarray}
P(C)&=&\mathcal{R}_q(m(q-1)-\nu_1-\nu_2^\perp-1,m)\nonumber\\
&=&\mathcal{R}_q(\nu_2-\nu_1,m)
\end{eqnarray}
By \cite[Theorem 11]{grassl03}, if there exists a vector of
weight $r$ in $P(C)$, then there exists an $[[r,k',d']]_q$ quantum code,
where $k'\geq (k(\nu_2)-k(\nu_1)-q^m+r)$ and distance $d'\geq d$.
Since $P(C) = \mathcal{R}_q(\nu_2-\nu_1,m) \supseteq
\mathcal{R}_q(\mu,m)$ for all $0\leq\mu\leq \nu_2-\nu_1$, the
weight distributions of $\mathcal{R}_q(\mu,m)$ give all the
lengths to which $Q$ can be punctured. Moreover $P(C)$
will certainly contain vectors whose weight $r=d(\mu)$, that is
the minimum weight of $\mathcal{R}_q(\mu,m)$. Thus there exist
punctured quantum codes with the parameters
$[[d(\mu),\geq (k(\nu_2)-k(\nu_1)-q^m+d(\mu)),\geq d]]_q$.
\end{proof}
\smallskip

It is also possible to puncture codes constructed via 
Theorem~\ref{th:grm_herm}. However, we need to redefine the 
puncture code. Recall a quantum code constructed 
via the Hermitian construction has an underlying classical code
$C\subseteq \F_{q^2}^n$. The (Hermitian) puncture code, $P_h(C)$, is defined as 
\cite[Theorem 13]{grassl03}
\begin{eqnarray}
P_h(C)=\{ \tr_{q^2/q}(a_ib_i^q)_{i=1}^n \mid a,b \in
C\}^\perp,\label{eq:puncdef2}
\end{eqnarray}
where the subscript is used to differentiate the two 
constructions. Again the nonzero weights in the puncture code
determine the possible puncturings of the quantum code. Now we
shall apply these ideas to the puncturing of quantum Reed-Muller
codes constructed using the Hermitian construction. 
\smallskip

\begin{theorem}
Let $C=\mathcal{R}_{q^2}(\nu,m)$ with $0\leq \nu\leq m(q-1)-1$ and
$ (q+1)\nu \leq \mu \leq m(q^2-1)-1$.  The puncture code
$P_h(C)\supseteq \mathcal{R}_{q^2}(\mu,m)^\perp|_{\F_q}$. If
$P_h(C)$ contains a vector of weight $r$, then there exists an
$[[r,\geq (r-2k(\nu)),\geq d(\nu^\perp)]]_q$  quantum code\footnote
{Since C is over $\F_{q^2}$, $q^2$ should be used in
equations (\ref{eq:grm_dim}) and (\ref{eq:grm_dist}).}.
\end{theorem}
\smallskip
\begin{proof}
By equation (\ref{eq:puncdef2}) the puncture code
$P_h(C)$ is given as 
\begin{eqnarray}
P_h(C)=\{ \tr_{q^2/q}(a_ib_i^q)_{i=1}^n \mid a,b \in C\}^\perp.
\end{eqnarray}
To get a more useful description of $P_h(C)$ let us consider
the code $D$ given by
\begin{eqnarray}
D=\{ (a_ib_i^q)_{i=1}^n \mid a,b \in C\}.
\end{eqnarray}
If $C=\mathcal{R}_{q^2}(\nu,m)$, then $a=ev$ $f$ and $b=ev$ $g$ for some 
$f,g$ in $L_m(\nu)$. Then $a_i b_i^q =(ev$ $fg^q)_i$ and we can write
\begin{eqnarray}
D&=& \{  ev\, fg^q\mid f,g \in  L_m(\nu) \}. \nonumber
\end{eqnarray}
Since deg $fg^q\leq (q+1)\nu$, $fg^q$ is in $L_m(q\nu+\nu)$ and it follows
\begin{eqnarray}
D&\subseteq &\{  ev\, f \mid f \in  L_m((q+1)\nu) \}, \nonumber \\
D& \subseteq & \RM_{q^2}(\mu,m),
\end{eqnarray}
for $(q+1)\nu\leq \mu \leq m(q^2-1)-1$. By Delsarte's theorem
\cite[Theorem 2]{delsarte75}, the dual of the trace code is the
restriction of the dual code. That is, 
\begin{eqnarray}
\tr_{q^2/q}(D)^\perp&=&D^\perp|_{\F_q} = D^\perp \cap \F_q^n.
\end{eqnarray}
However, as $P_h(C)^\perp =\tr_{q^2/q}(D)$, it follows that
\begin{eqnarray}
P_h(C) & =&  D^\perp|_{\F_q} \supseteq
\mathcal{R}_{q^2}(\mu,m)^\perp |_{\F_q},\\
 P_h(C) &\supseteq &\mathcal{R}_{q^2}(m(q^2-1)-\mu-1,m)|_{\F_q}
\end{eqnarray}
From Theorem \ref{th:grm_herm} there exists an
$[[q^{2m},q^{2m}-2k(\nu),d(\nu^\perp)]]_q$ quantum code $Q$ for
$0\leq \nu\leq m(q-1)-1$. By \cite[Theorem 11]{grassl03}, 
$Q$ can be punctured to all lengths which have nonzero
distribution in $P_h(C)$. Thus if $P_h(C) \supseteq
\mathcal{R}_{q^2}(\mu^\perp,m)|_{\F_q}$ contains a vector of
weight $r$, then there exists an $[[r,\geq (r-2k(\nu)),\geq
d(\nu^\perp)]]_q$ code.
\end{proof}
%

\section{Quantum MDS Codes via GRM Codes}\label{sec:mds}
MDS codes occupy a special place of interest in coding theory in view of their
optimality with respect to the Singleton bound and also because of their
many connections to other branches of mathematics. So in this section
we shall construct some linear quantum MDS codes as a special case of
quantum GRM codes. The usefulness of such an approach will be appreciated
later when we try to puncture them. 

\subsection{Quantum MDS codes}
The previous results enable us to derive very easily some 
qauntum MDS codes as a special case.
\smallskip

\begin{corollary}
There exist quantum MDS codes with the parameters
$[[q,q-2\nu-2,\nu+2]]_q$ for $0 \leq \nu \leq (q-2)/2$ and
$[[q^2,q^2-2\nu-2,\nu+2]]_q$ for $0 \leq \nu \leq q-2$.
\end{corollary}
\smallskip
\begin{proof}
These codes and the corresponding ranges of $\nu$ are a direct
consequence of Corollary~\ref{co:grm_css2} and Theorem~\ref{th:grm_herm} 
with $m=1$. In both cases it can be verified that
$k(\nu)=\nu+1$ and $d(\nu^\perp)=\nu+2$. The quantum codes
$[[q,q-2\nu-2,\nu+2]]_q$ and $[[q^2,q^2-2\nu-2,\nu+2]]_q$ follow
on substituting the values of $k(\nu)$ and $d(\nu^\perp)$
in these constructions. It can be easily verified that these
quantum codes satisfy the quantum Singleton bound \cite{rains99}.
\end{proof}
%
It must be noted that when applying equations (\ref{eq:grm_dim})
and (\ref{eq:grm_dist}) we must be careful to use $q$ for
Corollary~\ref{co:grm_css2} and $q^2$ for Lemma~\ref{th:grm_herm}.
An alternate approach to constructing these codes can be found in 
\cite{grassl03}.

\subsection{Puncturing Quantum MDS codes.}
It was shown in \cite{grassl03} that there exist $q$-ary quantum
MDS codes of lengths $3\leq n\leq q$ as well as $q^2-s$ for some
$s>1$. The range and the distribution of such $s$ was not answered
analytically. However numerical evidence was provided which
indicated that there existed many $s$ or equivalently many MDS
codes with lengths between  $q$ and $q^2$. Here, we partly address
this problem by analytically proving the existence of MDS codes of
length $(\nu+1)q$ with $0\leq \nu \leq q-2$.

We will assume that the quantum code $Q$ is constructed
using Theorem \ref{th:grm_herm} and the corresponding classical code
is $C$. As mentioned earlier, 
the associated puncture code $P_h(C)$ determines the lengths to which 
we can puncture $Q$ (refer Theorem
\ref{th:grm_punc2}, \cite[Theorem 3]{rains99} and \cite[Theorem
11,13]{grassl03}). In general it is not easy to calculate the weight
distribution of $P_h(C)$, but we can simplify our task by calculating
either the weight distribution or the minimum distance of a
subcode in $P_h(C)$. We will need the following result proved in
\cite{pellikaan04}, though it is not stated explicitly as a
theorem. We have restated it slightly differently to make the proofs clearer.
\smallskip

\begin{lemma}\label{th:grm_subcode}
Every GRM code $\mathcal{R}_{q}(\nu,m)$ is a subcode of
$\mathcal{R}_{q^m}(q^m-d(\nu),1)$. Alternatively,
\begin{equation}
\mathcal{R}_{q}(\nu,m) \subseteq
\mathcal{R}_{q^m}(q^m-d(\nu),1)|_{\F_q}.
\end{equation}
\end{lemma}
\smallskip

\begin{proof}
See \cite{pellikaan04}. The result in \cite{pellikaan04} actually
states that
\begin{equation}
\mathcal{R}_{q}(\nu,m) \subseteq
GRS_{q^m-d(\nu)+1}(\mbf{a},\mbf{1})|_{\F_q},
\end{equation}
where $\mbf{a}=(0, 1,\zeta, \zeta^2, \ldots, \zeta^{q^m-2})$,
$\zeta$ is a primitive element in $\F_{q^m}$ and $GRS_{q^m-d(\nu)+1}(\mbf{a},\mbf{1})$
stands for the generalized Reed-Solomon code with the parameters 
$[q^m,q^m-d(\nu)+1,d(\nu)]$ (see \cite[pages~175-178]{huffman03}). However, as
$\RM_{q^m}(q^m-d(\nu),1)$ is the extended Reed-Solomon code,
it is the same as $GRS_{q^m-d(\nu)+1}(\mbf{a},\mbf{1})$.
Therefore, we can conclude that 
$ \mathcal{R}_{q}(\nu,m) \subseteq
\mathcal{R}_{q^m}(q^m-d(\nu),1)|_{\F_q}$.
\end{proof}
\smallskip

\begin{lemma}\label{th:mds_punc}
Let $C=\mathcal{R}_{q^2}(\nu,1)$ with  $0\leq \nu \leq q-2$, then
the puncture code $P_h(C)$ has a vector of weight $(\nu+1)q$.
\end{lemma}
\smallskip

\begin{proof}
Let  $C=\mathcal{R}_{q^2}(\nu,1)$. By Theorem~\ref{th:grm_punc2}, we
know that $P_h(C) \supseteq \mathcal{R}_{q^2}(\mu,1)^\perp|_{\F_q}$,
where $ (q+1)\nu \leq \mu \leq (q^2-2)$.
\begin{eqnarray}
P_h(C) &\supseteq&\mathcal{R}_{q^2}(\mu,1)^\perp|_{\F_q}
=\mathcal{R}_{q^2}(q^2-\mu-2,1)|_{\F_q}
\end{eqnarray}
Choose $\mu=(\nu+1)q-2$. (Note that $(q+1)\nu \leq (\nu+1)q \leq
q^2-2$ holds for $0\leq\nu\leq q-2$). Then
\begin{eqnarray}
P_h(C)&\supseteq& \mathcal{R}_{q^2}(q^2-(\nu+1)q,1)|_{\F_q}.
\end{eqnarray}
We will show that $\mathcal{R}_q(q-\nu-1,2)$ is embedded in
$\mathcal{R}_{q^2}(q^2-(\nu+1)q,1)|_{\F_q}$; thus in $P_h(C)$ also. By
equation~(\ref{eq:grm_dist}), wt$(\mathcal{R}_q(q-\nu-1,2))=d(q-\nu-1)=(\nu+1)q$. By Lemma
\ref{th:grm_subcode} $\mathcal{R}_q(q-\nu-1,2)$ is embedded in
$\mathcal{R}_{q^2}(q^2-(\nu+1)q,1)|_{\F_q} \subseteq P_h(C)$. Hence
$P_h(C)$ contains a vector of weight equal to $(\nu+1)q$.
\end{proof}
\smallskip

\begin{theorem}
There exist quantum MDS codes with the parameters $[[(\nu+1)q,
(\nu+1)q-2\nu-2,\nu+2]]_q$ for $0\leq \nu\leq q-2$.
\end{theorem}
\smallskip

\begin{proof}
The puncture code of $C=\mathcal{R}_{q^2}(\nu,1)$ has a vector of
weight $(\nu+1)q$ by Lemma \ref{th:mds_punc}. By Theorem~\ref{th:grm_punc2},
 the quantum code $[[q^2,q^2-2\nu-2,\nu+2]]_q$ derived from $C$ can be
punctured to a length of $(\nu+1)q$, giving a code
$[[(\nu+1)q,(\nu+1)q-2\nu-2,\nu+2]]_q$ which can be verified to be a
quantum MDS code.
\end{proof}

\section{Conclusion}
We constructed a family of nonbinary quantum codes based on
the classical generalized Reed-Muller codes. Then we studied when
these codes can be punctured to give more quantum codes. As a
special case we derived a series of quantum MDS codes from the
generalized Reed-Muller codes. We provided a partial answer to the
question of existence of $q$-ary MDS codes with lengths in the
range $q$ to $q^2$ by analytically proving the existence of a
series of codes with lengths in this range.

\section*{Acknowledgment}
This research was supported by NSF CAREER award CCF~0347310,
NSF grant CCR 0218582, a Texas A\&M TITF initiative, and a TEES Select
Young Faculty award.

\bibliography{grm_refs}
%

\end{document}